# Nanomechanics of Shear Rate-Dependent Stiffening in Micellar Electrically Conductive Polymers

Jingchen Wang[1], Tianqi Hu[2], Jingjie Yeo[3,*]

[1]School of Materials Science and Engineering, Cornell University, Ithaca, NY, USA

[2]Manlius Pebble Hill School, Syracuse, NY, USA

[3]Sibley School of Mechanical and Aerospace Engineering, Cornell University, Ithaca, NY, USA

**Abstract**

Electrically conducting polymers with mechanical adaptability are essential for flexible electronics, yet most suffer structural degradation under rapid deformation. In this study, multiscale coarse-grained (MSCG) simulations are used to uncover the nanoscale origins of an unusual strain-rate-dependent stiffening in a poly(2-acrylamido-2-methyl-1-propanesulfonic acid) (PAMPSA)-polyaniline (PANI) blend. The self-assembled morphology consists of semi-crystalline PANI-rich micellar cores dispersed in a soft, viscoelastic PAMPSA matrix. At low shear rates, micelles migrate and coalesce into larger aggregates, enhancing local crystallinity and transient entanglement density while dissipating stress through matrix deformation. At high shear rates, micelles cannot reorganize quickly enough, leading to core dissociation and the emergence of highly aligned PANI filaments that directly bear the load, with PAMPSA serving as a weak but extended support phase. These contrasting regimes (densification-driven local alignment versus dissociation-driven global alignment) enable reversible mechanical stiffening across three orders of magnitude in shear rate. The results provide a molecular-level framework for designing solid-state polymers with tunable, rate-adaptive mechanical properties.

**Keywords**：Rate adaptive; Polymer; Coarse grain; Molecular dynamics.

# 1. Introduction

The development of next-generation materials capable of withstanding dynamic mechanical environments is of critical importance, particularly in the realm of soft electronics, where such materials underpin the advancement of wearable, flexible, and bio-integrated devices. Among potential candidates, electrically conducting polymers have attracted significant attention due to their unique combination of mechanical flexibility and electronic conductivity, making them indispensable for technologies that demand both structural adaptability and functional performance [1–6]. Despite these advantages, conducting polymers remain limited by their mechanical fragility under rapid or repeated deformation. Such weaknesses can lead to material failure in devices that must endure continuous or sudden mechanical stresses, including stretchable displays, conformable sensors, and bioelectronic interfaces [7–12]. Overcoming this challenge requires the development of strategies that enhance the mechanical resilience of conducting polymers without compromising their electrical functionality.

Recent studies have shown that certain polymer blends can exhibit unusual mechanical strengthening under high shear rates. A binary mixture of poly(2-acrylamido-2-methyl-1-propanesulfonic acid) (PAMPSA) and polyaniline (PANI), plasticized with propane sulfonic acid (PSA), displays pronounced strain-rate-dependent stiffening, reminiscent of the shear-thickening behavior observed in non-Newtonian fluids such as aqueous cornstarch suspensions [13]. This phenomenon is particularly compelling because the PANI-PAMPSA system is a solid-state material, yet it exhibits a stress-dependent increase in resistance to deformation typically associated with complex fluids. It is hypothesized that this behavior arises from the nanoscale morphology of the polymer network, specifically the formation of micellar aggregates and their organization into an interconnected structure. At low shear rates, these micellar domains undergo dynamic viscoelastic relaxation, whereas at high shear rates, the enhanced resistance to deformation arises from micelle core dissociation, which significantly reinforces the material's mechanical response and enables adaptive mechanical stiffening [13]. However, direct experimental characterization of these nanoscale features, particularly under deformation, remains a significant challenge. Conventional imaging techniques, including electron microscopy and atomic force microscopy, lack the temporal resolution and *in situ* capabilities necessary to capture the dynamic evolution of micellar networks under mechanical stress.

Considering these experimental limitations, molecular simulations provide a powerful complementary approach for probing the nanoscale structure and mechanics of the PANI–PAMPSA system. Among these, the multiscale coarse-graining (MSCG) methodology has emerged as a robust framework for studying self-assembling soft matter systems. By mapping groups of atoms into coarse-grained "beads," MSCG techniques substantially reduce computational complexity while preserving the essential physicochemical interactions [14–16]. This approach enables the simulation of mesoscale phenomena in systems that are prohibitively large for fully atomistic models. It is particularly well-suited for systems such as the PANI–PAMPSA mixture, which exhibit hierarchical self-assembly and dynamic crosslinking behaviors that are difficult to capture with traditional atomistic simulations [17–19].

In this study, we employ MSCG simulations to elucidate the nanoscale mechanisms underlying the strain-rate-adaptive mechanical properties of the PANI-PAMPSA polymer blend. We first examined the self-assembly of the PANI-PAMPSA system, confirming the formation of micellar structures through analyses of mean squared displacement (MSD), radial distribution functions (RDF), and density distribution maps. Subsequently, we investigated the shear deformation of the self-assembled system across different shear rates, characterizing the deformation mechanisms using stress–strain curves, binding energy profiles, and atomic stress distributions. By revealing the molecular-level mechanisms that govern strain-rate-dependent reinforcement, this study advances the broader objective of developing self-protective materials capable of maintaining functional integrity under demanding operational conditions.

## 2. Materials and Methods

### 2.1 Atomistic Model Construction and Equilibration

To investigate the self-assembly mechanisms and mechanical properties of the PAMPSA–PANI binary polymer system, the all-atom (AA) simulation of the PANI-PAMPSA mixture was first conducted to provide the basis for coarse-grained (CG) force field parameterization. Atomistic models were constructed using the PySIMM tool [20], with a PAMPSA:PANI molar ratio of 10:8 and a chain length ratio of 100:10, which agrees with the experimentally prepared samples [13].

All molecular dynamics simulations, including both AA and CG models, were performed using the Large-scale Atomic/Molecular Massively Parallel Simulator (LAMMPS) software package [21]. For each AA model, an initial equilibration was conducted in the canonical (NVT) ensemble using a 1 fs time step. The system was heated from 1 K to 300 K over 1 ns, followed by a 1 ns equilibration at 300 K. Subsequent simulations were carried out in the isothermal-isobaric (NPT) ensemble with 1 fs time step and 1 bar pressure, including heating from 300 K to 450 K for 1 ns, quenching from 450 K to 300 K for 1 ns, and final equilibration at 300 K for 1 ns. The final 1 ns trajectory, consisting of 100 evenly spaced frames, was used for CG parameter extraction.

**2.2 Coarse-Grained Simulation and Force Field Generation**

Coarse-grained models of the polymers were generated using the MSCG frameworks developed by Voth and co-workers. In this approach, groups of atoms corresponding to chemically distinct functional units were mapped onto CG "beads," each defined by its center-of-mass position and the net force acting on the constituent atoms. The coarse-graining scheme for each polymer type is illustrated in Fig. 1 (a), where the PANI monomer is treated as a single bead "ANI", and the PAMPSA acrylamide backbone as the bead "ACR", and functional group methylpropanesulfonic acid as the bead "PMSA".

Fig. 1 (b) demonstrates the workflow of the MSCG force field generation. CG trajectories were first generated by applying this mapping to the 100-frame AA trajectory of each system. The first CG frame was selected as the initial configuration for subsequent CG simulations. Force-matching methods using numerical interpolation of each atomic position and force with B-spline basis functions were employed to derive bonded, angular, and non-bonded potentials from the mapped CG data, ensuring that the coarse-grained forces reproduced those observed in the AA models. These CG potentials were implemented in LAMMPS to simulate the coarse-grained system under NPT conditions with 1 fs time step, 1 bar, and 300 K for 1 ns. The resulting CG trajectory was analyzed to compute RDFs for both bonded and non-bonded bead pairs. These structural descriptors were then compared against those obtained from the AA simulations to assess the fidelity and transferability of the derived CG force field. The final non-bonded pair, bond, and angle potentials obtained by the force matching algorithm is shown in Fig 1 (c).

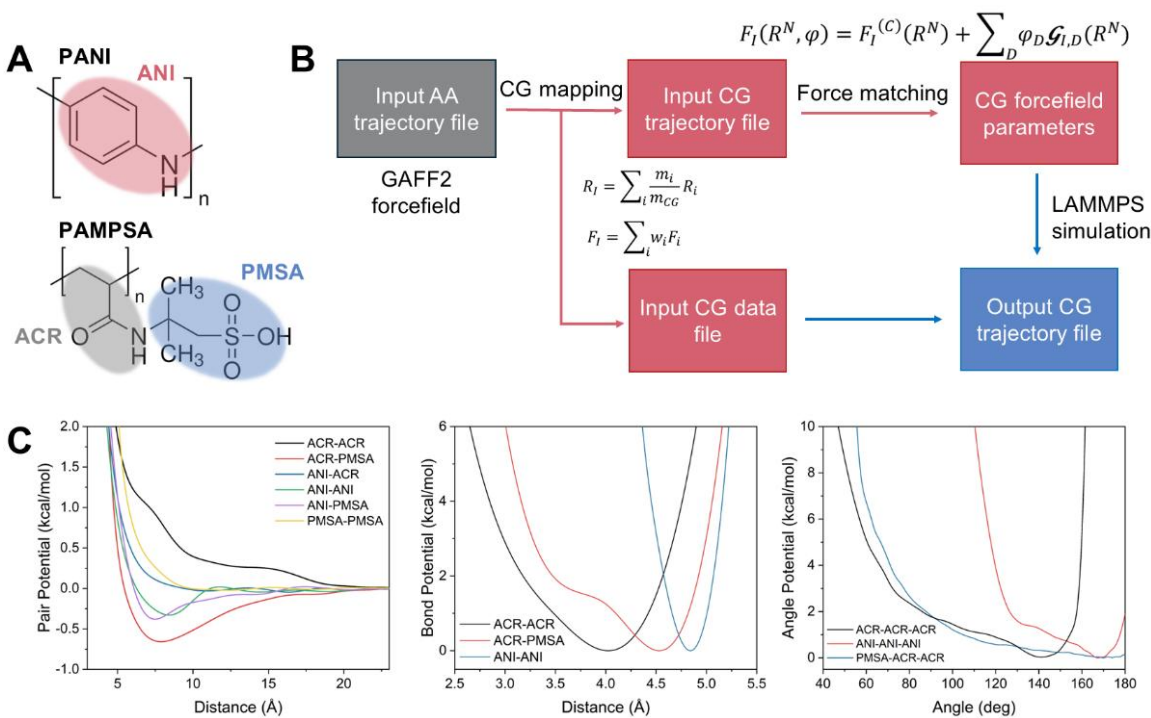

**Figure. 1.** Multiscale coarse-graining of the PANI-PAMPSA system. (a) Mapping of atomistic to coarse-grained beads of PANI and PAMPSA based on functional group identities. (b) A schematic workflow of the simulation pipeline, beginning with AA model generation and equilibration, followed by CG mapping, force-matching parameterization, and subsequent CG molecular dynamics simulations. (c) CG force field potentials for all bead types obtained by the force-matching procedure, including non-bonded pair, bond, and angle potentials.

## 2.3 Self-Assembly and Mechanical Deformation Analysis

After validating the CG force fields, the self-assembly and mechanical response under rapid deformation of the PANI-PAMSA mixture was investigated. The system was equilibrated under NPT conditions at 300 K and 1 atm, and spontaneous microstructural changes were allowed to proceed over nanosecond-scale simulation times. Mechanical deformation was imposed with simple shear applied in the xy, yz, and zx planes at various shear rates ($10^6$–$10^8$ s$^{-1}$). Shear stress–strain curves were computed from the off-diagonal components of the pressure tensor using the virial formulation. The resulting curves were used to extract key mechanical descriptors such as shear modulus, yield stress, and strain softening or hardening behavior. To quantify micellar network connectivity and chain entanglement, the shortest multiple disconnected path of the polymer chains were computed using the Z1+ algorithm [22]. In this approach, polymer chains are reduced to their primitive paths using a purely geometrical method. The algorithm shortens each chain's contour length as much as possible while keeping the chain ends fixed in their original positions and preventing any crossings between chains. This process produces a network of primitive paths that captures the underlying entanglement topology. From this network, the spatial locations of individual entanglement points can be directly identified, and the number of kinks along each chain can be counted. Because the method is geometrical rather than statistical, it does not require any prior assumptions about the conformational statistics of the chains, making it particularly well suited for analyzing anisotropic or deformed systems. For each polymer type, the squared end-to-end distance $<R^2>$, primitive path contour length $<L_{pp}>$, and mean number of kinks per chain $<Z>$ were evaluated before and after deformation. $<R^2>$ provides a measure of chain extension, $<L_{pp}>$ reflects the contracted primitive path length, and $<Z>$ characterizes the topological connectivity of the entanglement network. All quantities were averaged over the true polymer chains and reported as ensemble means for each deformation state.

To evaluate polymer alignment and orientation under shear, the bond angle distributions between polymer bond vectors and the shear direction were calculated at different strains. Specifically, for each bonded pair of coarse-grained beads, the angle θ between the bond vector and the non-shear axis was calculated. Distributions of θ were used to quantify polymer alignment during shear deformation. A shift toward θ ≈ 0 under increasing strain indicates preferential alignment along the shear direction, which is consistent with strain-induced ordering or domain reorientation.

## 3. Results and Discussion

### 3.1 Self-assembly and characterization of PANI-PAMPSA micelles

The self-assembly of the PANI-PAMPSA system is first studied. Fig. 2(a) shows the formation of nanoscale micellar aggregates of PANI in the PAMPSA matrix from the initially randomized PANI–PAMPSA blend within 10 ns. At 5 ns, the PANI chains form initial aggregations as they move through the PAMPSA matrix, which then begin to condense into micelle cores observed at 10 ns. Fig. 2 (b) shows the density distribution maps of PANI and PAMPSA in the initial and final configurations, which confirm that the PANI-rich segments cluster into distinct cores, while the sulfonated PAMPSA chains occupy the surrounding matrix, forming a core–shell morphology consistent with experimentally observed morphology of this system [13]. Notably, the computed radial distribution functions of the equilibrated structure in Fig. 2 (c) reveal relatively high intensity peaks for correlations between the PANI monomers, indicative of semi-crystalline packing in the micellar cores. Whereas the RDFs between PAMPSA beads are broad and relatively featureless, which is typical of a liquid-like polymer matrix. The PANI-PAMPSA cross RDF shows relatively low peaks, reflecting some interfacial interpenetration at the core–shell boundary. This suggests the self-assembly of PANI-PAMPSA into an interconnected core–shell micellar network with different nanostructures between the core and the matrix, which is consistent with prior experimental observations of the system [13].

The MSD data in Fig. 2 (d) indicate that the self-assembly dynamics involves two distinct stages, which are marked by changes in both the overall magnitude and the directional standard deviation of the MSD. In the initial regime, corresponding to the micellization onset, PANI chains display a relatively steep MSD slope and low standard deviation across the x, y, and z directions. This indicates rapid, isotropic diffusion as the chains are still dispersed and freely mobile throughout the system. Notably, PANI exhibits a significantly higher MSD than PAMPSA at all timescales, reflecting its higher mobility and lower degree of initial entanglement or steric hindrance compared to the more extended and larger PAMPSA chains. During the second regime, the micellar clusters of PANI begin to form and condense as the MSD of PANI decreases in slope, indicating reduced mobility as chains become confined within micelle cores. In addition, the standard deviation among the directional MSDs increases, signaling the onset of anisotropic motion. This results from dynamic heterogeneity introduced by micelle formation, where some PANI chains become trapped in dense aggregates, while others that are particularly at the micelle periphery or within more mobile clusters retain partial freedom of movement. Additionally, drift or reorganization of entire micelles can contribute to uneven displacements in different directions, further elevating the MSD standard deviation. This two-regime behavior highlights the transition from free diffusion to structurally constrained motion and supports the conclusion that micellization is driven by PANI self-aggregation within a relatively immobile PAMPSA matrix. Therefore, the density profiles, RDFs, and MSDs indicate a well-ordered PANI core surrounded by a disordered, mobile PAMPSA shell.

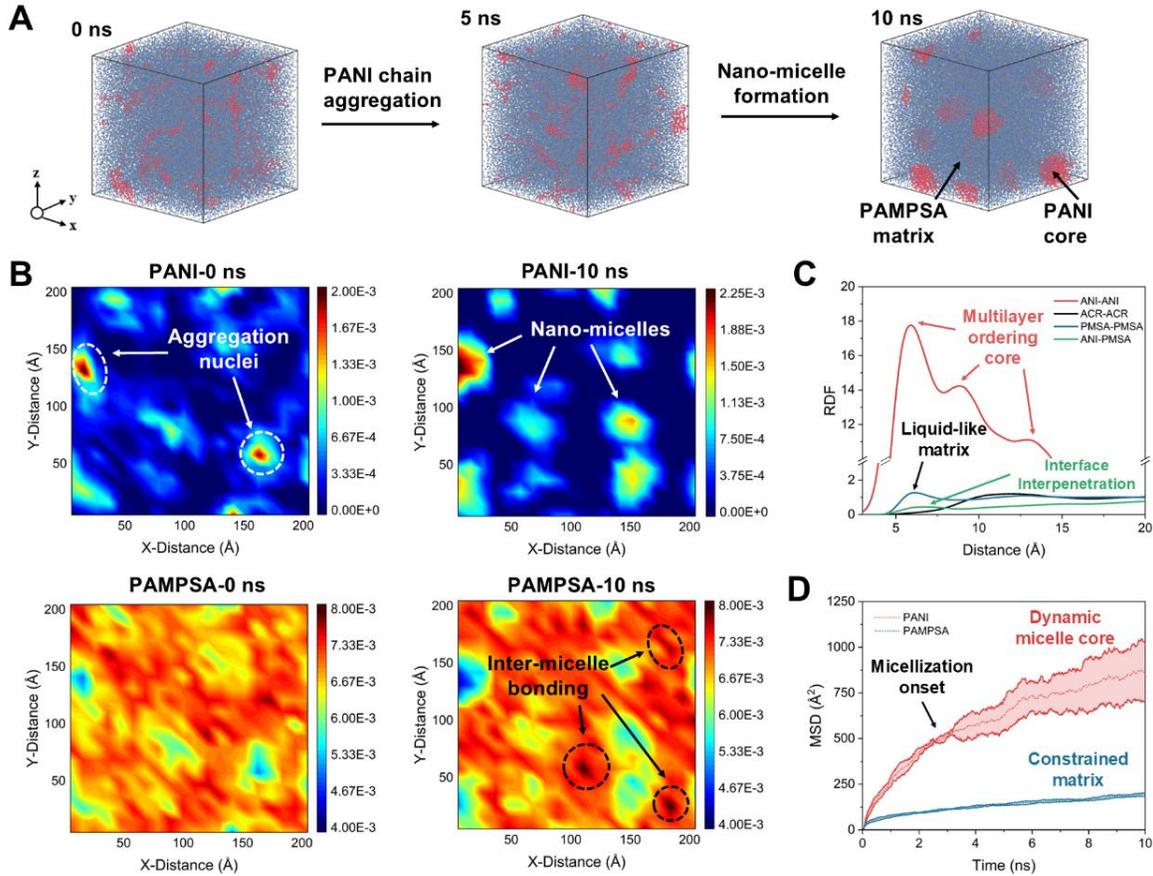

**Figure 2.** Self-assembly and structural characterization of PANI-PAMPSA mixtures. (a) Snapshots of spontaneous micelle formation from randomized initial configurations at 0ns, 5ns, and 10 ns. (b) Density distribution snapshots of the initial and final configuration of PANI and PAMPSA show aggregation of PANI as the nano-micelle core and PAMPSA as the inter-micellar matrix. (c) RDFs between key bead types in the final equilibrated structure indicate semi-crystalline ordering in the micelle core, a liquid-like matrix, and interfacial penetration between core and matrix. (d) MSD shows relatively high mobility of the micelle core within relatively more constrained matrix after the micellization onset.

**3.2 Deformation response of PANI-PAMPSA at different shear rates**

To evaluate the shear deformation response, the mechanical behavior of the self-assembled PANI–PAMPSA system was characterized under shear deformation across a range of shear rates, from $10^7$ s$^{-1}$ to $10^{10}$ s$^{-1}$. Fig. 3(a) presents the stress–strain curves for shear deformation across different shear rates, from which the key mechanical properties were extracted and are summarized in Table 1. These quantitative results show that the shear modulus increases nearly 45-fold, from 1.07 MPa at $10^7$ s$^{-1}$ to 48.13 MPa at $10^{10}$ s$^{-1}$, reflecting a dramatic stiffening of the network at higher rates. Similarly, the yield stress rises from only 0.03 MPa at $10^7$ s$^{-1}$ to 4.85 MPa at $10^{10}$ s$^{-1}$, which experienced an increase of over two orders of magnitude. The strain at yield also grows with rate, from 3.07% to 10.21%, indicating that the network can sustain more deformation before yielding at faster rates, which agrees with the experimentally observed response [13]. The ultimate tensile strength (UTS) shows an even steeper rate-dependence, increasing from 0.29 MPa at the slowest rate to 16.11 MPa at the fastest, consistent with the steep post-yield strain-hardening observed in the curves. Together, these trends confirm the rate-dependent stiffening behavior of PANI-PAMPSA. At low shear rates, the network deforms more easily, while at high shear rates, the material exhibits marked shear thickening and substantially greater load-bearing capacity.

**Table 1.** Summary of mechanical properties under shear and tensile deformation. Key mechanical parameters derived from stress–strain measurements, including elastic modulus, yield stress, maximum stress, and failure strain, are tabulated for shear deformation (xy, xz, yz) and uniaxial tension (x, y, z) across polymer compositions.

|  | Shear modulus (MPa) | Yield stress (MPa) | Strain at yield (%) | UTS (MPa) |
|---|---|---|---|---|
| $\dot{\varepsilon} = 10^7$ | 1.07 | 0.03 | 3.07 | 0.29 |
| $\dot\varepsilon=10^8$ | 9.93 | 0.39 | 4.07 | 1.24 |
| $\dot\varepsilon=10^9$ | 18.92 | 1.85 | 9.86 | 5.26 |
| $\dot\varepsilon=10^{10}$ | 48.13 | 4.85 | 10.21 | 16.11 |

Fig. 3(b) illustrates the evolution of the total interaction energy between PANI and PAMPSA chains as a function of strain, revealing distinct deformation modes that vary with shear rate. At a shear rate of $10^7$ s$^{-1}$, the binding energy decreases gradually in magnitude through a series of discrete steps. These stepwise decreases correspond to the integration and merging of smaller micelles, which reduces the total surface area of PANI–PAMPSA contacts. Because PAMPSA can also incorporate into PANI micelles to form lamellar nanostructures [13], the micelles become denser in PANI as the interpenetrating PAMPSA chains lose contact and are pulled out under shear. As the shear rate increases, the stepwise micelle integration becomes less distinct. This behavior can be attributed to the micelles becoming increasingly elongated and structurally loosened at higher strain rates, which inhibits further integration. At the highest shear rate of $10^{10}$ s$^{-1}$, the binding energy increases in magnitude with strain, indicating micelle disintegration due to excessive shear, as PANI disperses into the PAMPSA matrix and chain contact intensifies.

The micellar structures in the final strained configuration are characterized by the monomer-monomer RDFs of PANI, as shown in Fig. 3(c). At a shear rate of $10^7$ s$^{-1}$, the RDF exhibits distinct peaks at 5.55 Å, 9.26 Å, 13.43 Å, and 17.62 Å, corresponding to short- and intermediate-range ordering within semi-crystalline micellar cores. As the shear rate increases, the peak intensities diminish, indicating reduced structural order in the micelles. Notably, the peak intensities are higher than those in the initial configuration at shear rates of $10^7$ s$^{-1}$ and $10^8$ s$^{-1}$, and remain comparable at $10^9$ s$^{-1}$, suggesting that micelle integration enhances the local ordering of PANI. This effect arises from shear-assisted chain alignment within the micelle cores, where deformation proceeds slowly enough to prevent micelle breakup yet allows polymer chains to rotate and slide into more favorable stacking arrangements. The process also relieves local packing defects formed during self-assembly, while micelle coalescence at these moderate shear rates increases local density and promotes tighter chain packing. At the highest shear rate of $10^{10}$ s$^{-1}$, the peak magnitudes decrease significantly, and the longer-range peaks disappear, indicating a pronounced loss of core order.

Fig. 3(d) presents snapshots of the PANI density distribution at strains of 100%, 200%, and 400% for shear rates of $10^7$, $10^8$, $10^9$, and $10^{10}$ s$^{-1}$. At $10^7$ s$^{-1}$, the micelles remain compact and approximately spherical, with diameters of about 3–5 nm. As strain increases, these micelles gradually integrate and merge into larger, denser aggregates up to 8 nm in size. At $10^8$ s$^{-1}$, micelle integration persists during the early stages, accompanied by micelle elongation and partial dissociation of the outer layers from the core. This results in a coexistence of dense, integrated micelles and more diffuse, loosely structured ones. At $10^9$ s$^{-1}$, aggregation continues, but the micelles become increasingly deformed and less dense. During this stage, filamentous PANI strands begin to separate from the micelles and align along the shear direction, extending over lengths of approximately 3–5 nm. At the highest shear rate of $10^{10}$ s$^{-1}$, PANI strands dissociate early, and the micelles disassemble into extended, aligned networks with pronounced directional orientation and low aggregation.

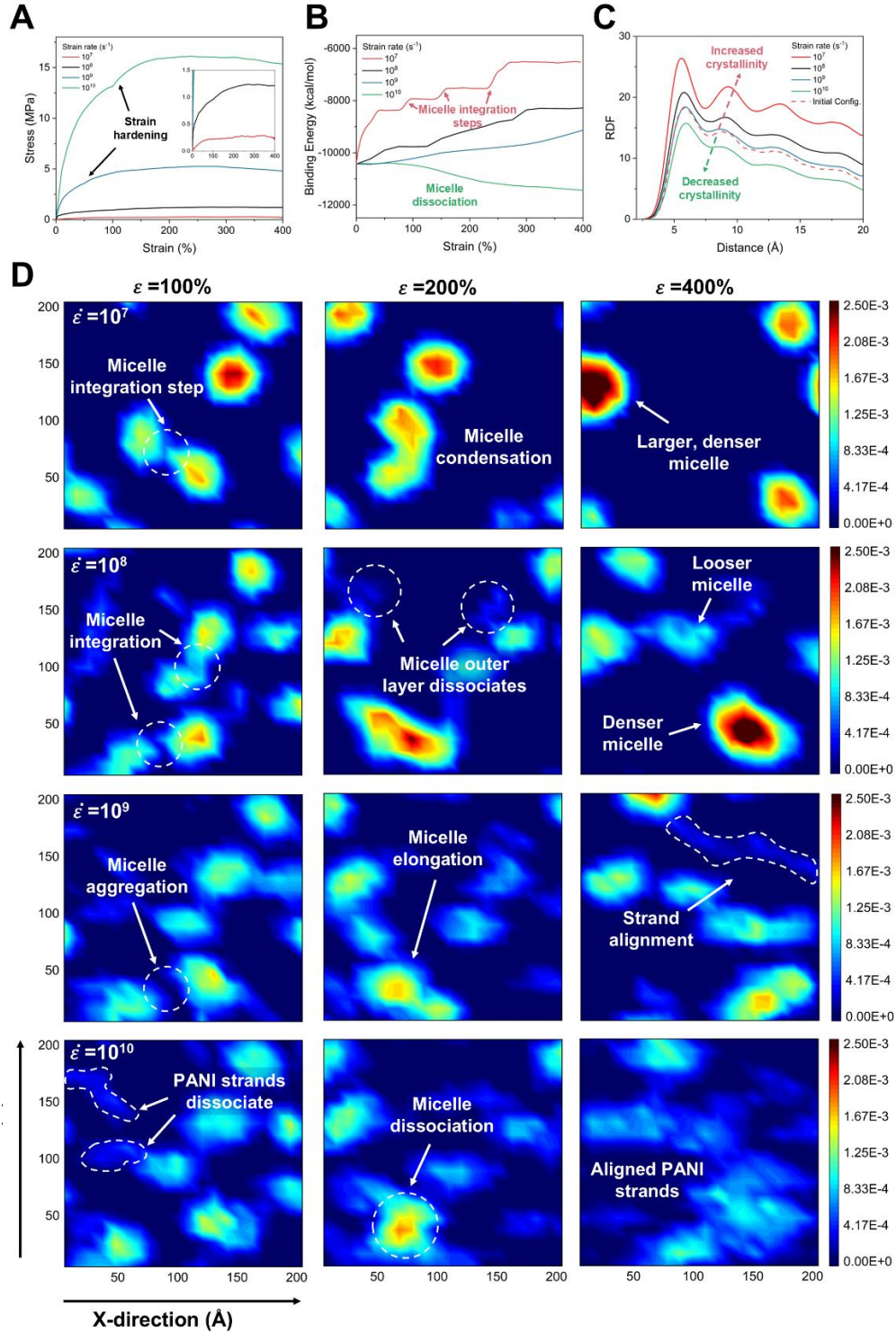

**Figure 3.** Mechanical properties and deformation response of PANI–PAMPSA under varying shear rates from $\dot{\varepsilon} = 10^7$ to $10^{10}$ s$^{-1}$. (a) Stress–strain curves at different shear rates reveal shear thickening behavior, with yield stress increasing at higher rates. (b) Evolution of the binding energy between PANI and PAMPSA with strain shows a transition in deformation mechanisms as shear rate increases. (c) RDFs between PANI monomers in the final strained configurations indicate that semi-crystalline micellar structures are more pronounced at lower shear rates and progressively diminish at higher rates. (d) Snapshots of PANI density distributions at strains of $\varepsilon = 100\%$, 200%, and 400% across different shear rates highlight the evolving mechanisms of energy dissipation and structural reorganization.

This sequence of transformations (from micelle coalescence to the formation of loose, aligned networks) illustrates the strain-rate adaptive behavior of the PANI–PAMPSA system, in which low-rate deformation involves viscoelastic micellar interactions, while high-rate deformation induces structural breakdown and the formation of stiff, load-bearing network structures.

### 3.3 Deformation mechanism of PANI-PAMPSA at different shear rates

Fig. 4(a) presents spatial stress distribution maps for PAMPSA and PANI at shear rates of $10^7$ and $10^{10}$ s$^{-1}$, at strains of 100%, 200%, and 400%. In all cases, PANI carries considerably higher stress than the surrounding PAMPSA matrix, confirming its role as the principal load-bearing component within the micellar network. At $10^7$ s$^{-1}$, stress in PANI is localized at micelle junctions during integration, particularly at 200% strain where two micelles merge. Once larger micelles form at 400% strain, stress becomes more evenly distributed over the merged domains, resulting in lower local stresses. The PAMPSA matrix at this low shear rate exhibits a diffuse, low-magnitude stress field, indicating a primarily supportive, non-load-bearing function. At the higher shear rate of $10^{10}$ s$^{-1}$, stress localization within PANI is more intense and sustained. At 200% strain, high stress accumulates at dissociating micelles, signaling the onset of structural breakdown. By 400% strain, PANI forms extended, aligned filaments that bear most of the applied load, as evidenced by concentrated stress along these elongated structures. Although the PAMPSA matrix still carries relatively low stress compared with PANI, its stress magnitude increases with shear rate. Overall, the stress maps reveal a rate-dependent load-bearing mechanism in which low shear rates promote transient stress localization at micelle junctions during integration, whereas high shear rates lead to direct stress transfer through aligned PANI chains following micelle dissociation.

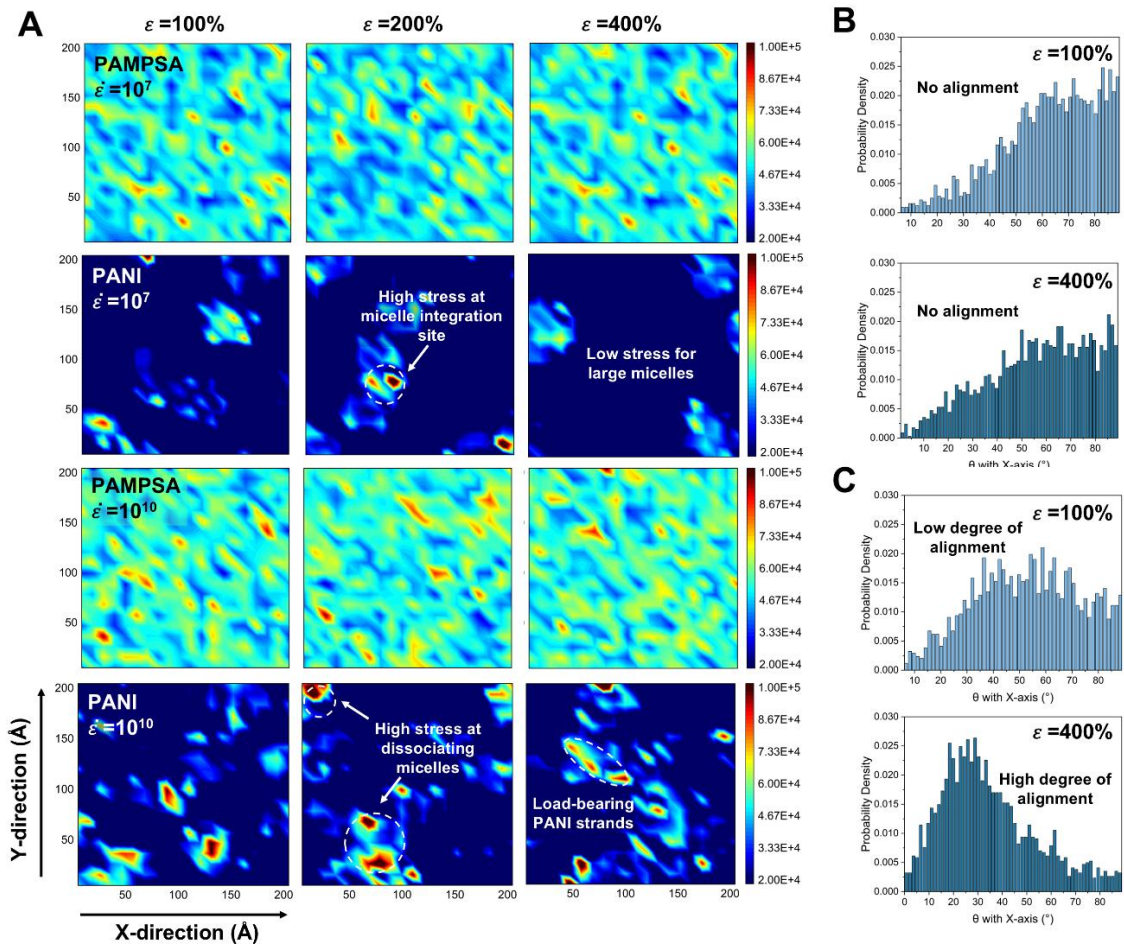

**Figure 4.** Investigation of the deformation mechanisms of PANI-PAMPSA under different shear rates. (a) Snapshots of the PANI and PAMPSA density distributions at strains of $\varepsilon = 100\%$, 200%,

and 400% for shear rates of $\dot{\varepsilon} = 10^7$ and $10^{10}$ s$^{-1}$ illustrate distinct load-bearing behaviors within the PAMPSA matrix and PANI phase, as well as rate-dependent changes in their respective load-sharing characteristics. (b) Bond angle distributions between the X-axis and PANI monomer bonds at the initial and final timesteps during shearing at $\dot{\varepsilon} = 10^7$ s$^{-1}$. (c) Bond angle distributions between the X-axis and PANI monomer bonds at the initial and final timesteps during shearing at $\dot{\varepsilon} = 10^{10}$ s$^{-1}$.

Fig. 4(b) and (c) present the bond angle distributions of PANI chains at strains of 100% and 400% for shear rates of $10^7$ and $10^{10}$ s$^{-1}$, respectively. At $10^7$ s$^{-1}$, the distributions remain broad and isotropic at both strain levels, indicating the absence of significant chain alignment under low-rate shear. In contrast, at $10^{10}$ s$^{-1}$, the distribution at 100% strain exhibits slight alignment, which becomes pronounced at 400% strain, where a distinct peak near 20° appears. This pronounced peak signifies strong chain alignment along the shear direction. The progression from isotropic to highly aligned configurations corresponds directly with the stress distribution results, demonstrating that enhanced load-bearing capability at high shear rates arises from the formation of a globally aligned PANI network.

The entanglement statistics summarized in Table 2 further substantiate the rate-dependent deformation mechanism. For PANI, both $<R^2>$ and $<L_{pp}>$ remain nearly constant, showing only slight increases across all shear rates. This indicates that the overall chain contour and primitive path lengths are largely preserved even at large strains. Notably, $<Z>$ increases slightly at $10^7$ s$^{-1}$ before decreasing at higher rates. The initial rise reflects the formation of transient entanglements during micelle coalescence, where increased micelle integration temporarily enhances chain–chain contact between PANI and PAMPSA. At higher shear rates, $<Z>$ for PANI decreases progressively, reaching 0.25 at $10^{10}$ s$^{-1}$, demonstrating that rapid deformation facilitates disentanglement and enables the high degree of chain alignment observed in Fig. 4(c) while maintaining load-bearing capacity through direct chain stretching.

In contrast, PAMPSA shows substantial increases in both $<R^2>$ and $<L_{pp}>$ across all shear rates, consistent with its role as the softer, more deformable phase that accommodates structural reorganization within the stiffer PANI network. As shear rate increases, $<R^2>$ continues to rise whereas $<L_{pp}>$ decreases, reflecting chain alignment and the loss of kinks. This results in greater chain extension in space but shorter, more direct primitive paths due to reduced entanglement. At $10^7$ and $10^8$ s$^{-1}$, $<Z>$ is significantly higher than in the initial state, indicating transient jamming and chain interlocking in the PAMPSA matrix as micelles reorganize, temporarily increasing entanglement density before complete network breakdown at higher rates. At $10^9$ and $10^{10}$ s$^{-1}$, PAMPSA chains remain extended yet weakly entangled, consistent with their function as an elongated, low-connectivity matrix surrounding the highly aligned PANI filaments. This dual alignment behavior underlies the strain-rate-adaptive mechanical response of the PANI-PAMPSA network.

**Table 2.** Average end-to-end distance $<R^2>$, mean contour length $<L_{pp}>$, and average Z-number $<Z>$ values for PAMPSA and PANI quantifying polymer chain entanglement before and after deformation.

|  | PANI- $<R^2>$ (Å) | PANI- $<L_{pp}>$ (Å) | PANI- $<Z>$ | PAMPSA- $<R^2>$ (Å) | PAMPSA- $<L_{pp}>$ (Å) | PAMPSA- $<Z>$ |
|---|---|---|---|---|---|---|
| Initial Config. | 38.05 | 38.24 | 0.48 | 44.92 | 43.29 | 0.30 |
| $\dot{\varepsilon}=10^7$ | 38.97 | 39.08 | 0.52 | 81.65 | 86.07 | 1.28 |
| $\dot{\varepsilon}=10^8$ | 38.68 | 38.70 | 0.41 | 85.76 | 82.94 | 0.85 |
| $\dot\varepsilon=10^9$ | 39.04 | 39.05 | 0.30 | 89.23 | 80.75 | 0.35 |
| $\dot\varepsilon=10^{10}$ | 38.66 | 38.57 | 0.25 | 87.22 | 78.84 | 0.33 |

## 3.4 Strain-rate adaptive mechanism of PANI-PAMPSA

The deformation response of the PANI–PAMPSA system arises from the dynamic restructuring of its micellar network, where stiff, semi-crystalline PANI-rich cores are embedded within a soft, deformable PAMPSA matrix. At equilibrium, the material consists of discrete micelles with limited interconnection. The applied shear rate controls the interplay among micelle coalescence, chain alignment, and entanglement rearrangement, leading to distinct deformation pathways. Fig. 5 summarizes three regimes and their corresponding rate-adaptive mechanisms. At low shear rates, micelles have sufficient time to migrate, merge, and condense into larger, denser aggregates with higher crystallinity. This coalescence process increases the entanglement density in both PANI and PAMPSA by bringing chains into closer contact. The applied load is transiently localized at micelle junctions during integration, while the PAMPSA matrix elongates and entangles to dissipate energy. The absence of significant global chain alignment indicates that deformation is primarily accommodated through micelle rearrangement and matrix flow rather than backbone stretching.

At intermediate shear rates, micelle association and elongation dominate. Stress is more evenly distributed between the PANI and PAMPSA phases compared with the low-rate regime, and the onset of directional alignment in the PANI phase begins to emerge. At high shear rates, micelles lack sufficient time to reorganize and begin to disintegrate. The PANI-rich cores unravel into extended, highly aligned chains that directly bear the applied load, while the softer PAMPSA matrix forms an elongated yet weakly connected network.

Thus, with increasing shear rate, the deformation mode transitions from densification-driven local alignment to dissociation-driven global alignment, defining the strain-rate–adaptive mechanical behavior of the PANI–PAMPSA network.

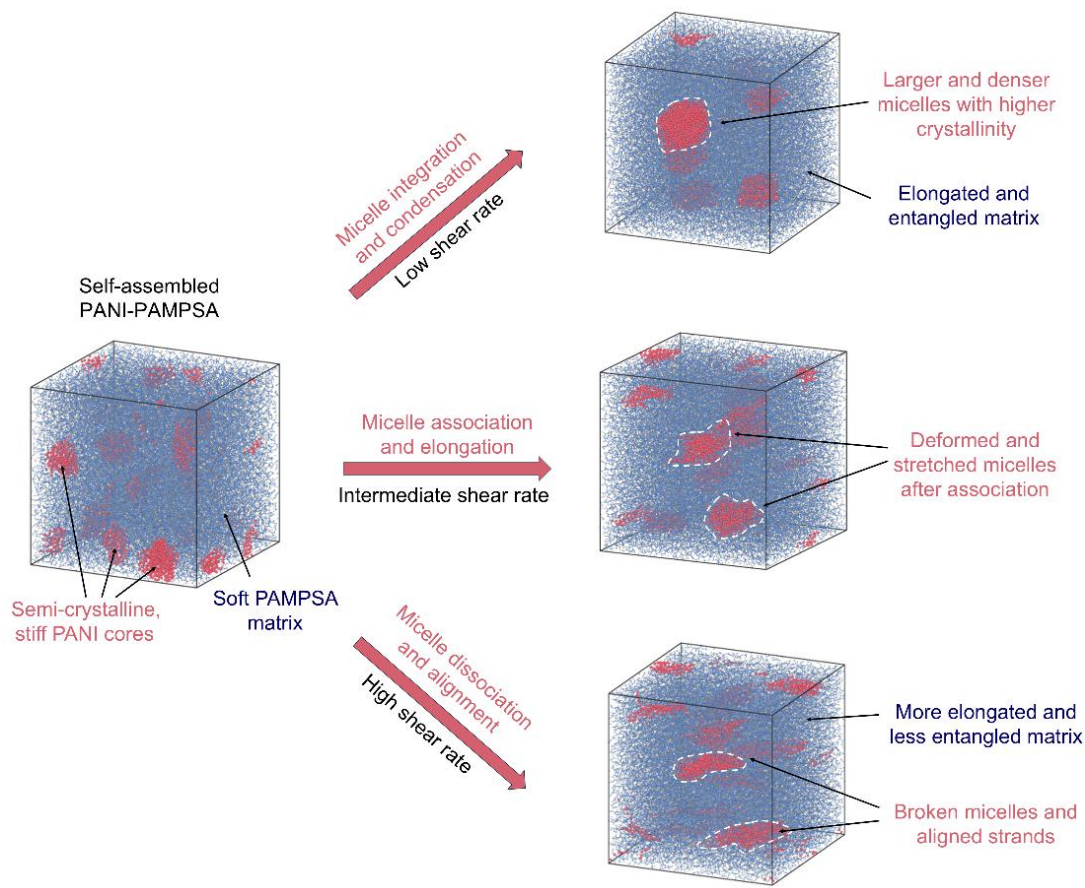

**Figure 5.** Schematic illustration of the rate-adaptive mechanism of PANI-PAMPSA mixture.

## 4. Conclusions

Multiscale coarse-grained simulations, combined with structural analyses including density mapping, radial distribution functions, and mean-squared displacement, revealed that the PANI-PAMPSA blend self-assembles into a core–shell micellar network comprising semi-crystalline PANI cores embedded in a viscoelastic PAMPSA matrix.

Shear deformation simulations across rates from $10^7$ to $10^{10}$ s$^{-1}$ demonstrated pronounced rate-dependent stiffening, evidenced by substantial increases in shear modulus, yield stress, and ultimate tensile strength. Binding energy evolution showed that low shear rates promote micelle migration and coalescence, whereas high rates induce micelle disintegration and enhanced PANI-PAMPSA interfacial contact.

Stress distribution and bond angle analyses confirmed that PANI acts as the primary load-bearing phase: at low rates, stress is localized at micelle junctions during integration, while at high rates, deformation produces globally aligned PANI filaments that directly carry the load. Entanglement statistics further supported a transition from transiently increased connectivity during micelle merging to reduced entanglement in the aligned high-rate regimes.

Together, these results establish that the shear-rate adaptability of PANI-PAMPSA arises from a transition between densification-driven local alignment and dissociation-driven global alignment, enabling reversible mechanical stiffening across several orders of magnitude in shear rate. These molecular-level insights provide a design framework for engineering conductive polymers with tunable mechanical resilience for flexible electronics and other mechanically dynamic applications.

While this model captures the fundamental mechanisms governing rate-dependent deformation, it does not fully represent fracture initiation and propagation, which are critical for predicting material failure under extreme conditions. Future work should extend this framework by incorporating fracture mechanics and explicit bond-breaking dynamics to better describe irreversible deformation. Additionally, introducing new molecular components or co-polymers into the blend could reveal synergistic strengthening effects and further enhance the tunability of both mechanical and electronic responses. Such studies would deepen understanding of how nanoscale structure and composition collectively govern macroscopic resilience in conductive polymer systems.

**Acknowledgments**

J.Y. acknowledges the support from the US National Science Foundation (NSF) under award CMMI-2338518. This work used SDSC Expanse at the San Diego Supercomputer Center through allocation BIO240093 from the Advanced Cyberinfrastructure Coordination Ecosystem: Services & Support (ACCESS) program supported by NSF grants #2138259, #2138286, #2138307, #2137603, and #2138296. D.W. acknowledges the support of the Dean Archer Undergraduate Research Program from Cornell University's Engineering Learning Initiative.

**CRediT author statement**

**Jingchen Wang:** Conceptualization, Methodology, Software, Data curation, Visualization, Investigation, Validation, Writing- Original draft preparation, Writing- Reviewing and Editing. **Tianqi Hu**: Software, Writing- Reviewing and Editing. ***Jingjie Yeo***: Conceptualization, Methodology, Supervision, Funding acquisition, Project administration, Writing- Reviewing and Editing.